\documentclass[bibyear]{aa} 
\usepackage{epsfig}
\usepackage{amsmath}
\usepackage{subfigure}
\usepackage{natbib}
\usepackage{multirow}
\usepackage{color}
\usepackage{longtable}
\usepackage{graphicx}
\usepackage{epstopdf}
\usepackage{booktabs}
\usepackage{txfonts}

\newcommand{\jmst}{J.~Mol.~Struct.}   
\newcommand{\chemrev}{Chem.~Rev.}
\newcommand{\jpcrd}{J.~Phys.~Chem.~Ref.~Data}

\begin{document}

\title{Tentative detection of HC$_5$NH$^+$ in TMC-1\thanks{Based on observations carried out with the Yebes 40m telescope (projects 19A003
and 20A014). The 40m radiotelescope at Yebes Observatory is operated by the Spanish Geographic Institute
  (IGN, Ministerio de Transportes, Movilidad y Agenda Urbana).}}

\author{
N.~Marcelino\inst{1},
M.~Ag\'undez\inst{1},
B.~Tercero\inst{2,3},
C.~Cabezas\inst{1},
C.~Berm\'udez\inst{1},
J.~D.~Gallego\inst{2},
P.~de Vicente\inst{2},
J.~Cernicharo\inst{1}
}

\institute{Grupo de Astrof\'isica Molecular, Instituto de F\'isica Fundamental (IFF-CSIC), C/ Serrano 121, 28006 Madrid, Spain.
\email: jose.cernicharo@csic.es
\and Centro de Desarrollos Tecnol\'ogicos, Observatorio de Yebes (IGN), 19141 Yebes, Guadalajara, Spain.
\and Observatorio Astron\'omico Nacional (IGN), C/ Alfonso XII, 3, 28014, Madrid, Spain.
}

\date{Received; accepted}

\abstract{Using the Yebes 40m radio telescope, we report the detection of a series
of seven lines harmonically related with a rotational constant $B_0$=1295.81581 $\pm$ 0.00026 MHz
and a distortion constant $D_0$=27.3 $\pm$ 0.5 Hz towards the cold dense cloud TMC-1.
Ab initio calculations indicate that the best possible
candidates are the cations HC$_5$NH$^+$ and NC$_4$NH$^+$.
From a comparison between calculated and observed rotational constants and other arguments based 
on proton affinities and dipole moments, we conclude that the
best candidate for a carrier of the observed lines is the protonated cyanodiacetylene cation, 
HC$_5$NH$^+$. The HC$_5$N/HC$_5$NH$^+$ ratio derived in TMC-1 is 240, which is very similar to 
the HC$_3$N/HC$_3$NH$^+$ ratio. Results are discussed in the framework of a chemical 
model for protonated molecules in cold dense clouds.
}


\titlerunning{HC$_5$NH$^+$ in TMC-1}
\authorrunning{Marcelino et al.}

\maketitle

\section{Introduction}
Although gas phase chemistry in cold interstellar clouds is dominated by
ion-neutral reactions, only around 15\% of the detected species are cations (see
the Cologne Database for Molecular Spectroscopy; \citealt{Muller2005}).
Observed polyatomic cations are usually protonated forms of stable molecules. The rather reduced
number of them detected in space is due to the fact that a dissociative recombination with electrons 
is rapid and depletes cations, producing most of
the neutrals observed in these objects. In addition to the widespread ions HCO$^+$ and
N$_2$H$^+$, other interesting protonated species are HCS$^+$ \citep{Thaddeus1981},
HCNH$^+$ \citep{Schilke1991},
HC$_3$NH$^+$ \citep{Kawaguchi1994},
HCO$_2^+$ (Turner et al. 1999; Sakai et al. 2008),
NH$_3$D$^+$ \citep{Cernicharo2013},
NCCNH$^+$ \citep{Agundez2015}, H$_2$COH$^+$ \citep{Bacmann2016}, and H$_2$NCO$^+$ \citep{Marcelino2018}.

The abundance ratio between a protonated molecule and its neutral counterpart, [MH$^+$]/[M], is
sensitive to the degree of ionisation and to the proton affinity of the neutral. The higher the 
density is, the lower the ionisation fraction is and thus the lower the importance of protonated molecules. It is
interesting to note that both the chemical models and the observations suggest a trend in which the 
abundance ratio [MH$^+$]/[M] increases with an increasing proton affinity of M \citep{Agundez2015}.

Protonated nitriles and dinitriles are observed in cold dense clouds because their neutral counterparts
are abundant and have high proton affinities. The nitriles HCN and HC$_3$N have proton
affinities of 712.9 kJ mol$^{-1}$ and 751.2 kJ mol$^{-1}$, respectively \citep{Hunter1998}, and abundances in
excess of 10$^{-8}$ relative to H$_2$ \citep{Agundez2013}. Whereas the dinitrile NCCN has a proton affinity
of 674.7 kJ mol$^{-1}$ \citep{Hunter1998} and an inferred abundance as large as that of HCN
and HC$_3$N \citep{Agundez2015,Agundez2018}. The next larger members in the series of cyanopolyynes and
dicyanopolyynes are also good candidates to be detected in their protonated form. Cyanodiacetylene (HC$_5$N)
has a high proton affinity (770 $\pm$ 20 kJ mol$^{-1}$; \citealt{Edwards2009}) and it is just a few
times less abundant than HC$_3$N in cold dark cores \citep{Agundez2013}. Dicyanoacetylene (NC$_4$N) also has a high proton
affinity (736 kJ mol$^{-1}$; this work) and it is likely to be present with a high abundance based on the large inferred abundance of NCCN. Hence, we could
expect the protonated forms of HC$_5$N and NC$_4$N to be present in cold dense clouds with moderately
high abundances.

In this Letter, we report the detection of a new series of harmonically related lines belonging to a
molecule with a $^1\Sigma$ ground electronic state towards the cold dark core TMC-1. Two of the molecular
species discussed above, HC$_5$NH$^+$ and NC$_4$NH$^+$, could be the carriers. From ab initio
calculations and the expected
intensities of the lines for each of these species, we conclude that we have discovered the cation
HC$_5$NH$^+$. No laboratory data are available for it, hence, this is the first time this species has been
observed. Abundance ratios between HC$_3$N and HC$_5$N and their protonated forms were derived and
compared with predictions from chemical models. We searched for lines that could be attributed to
NC$_4$NH$^+$ without success. A search for NC$_3$NC in our data also provides upper limits to the abundance of this species.

\section{Observations}

New receivers, which were built
within the Nanocosmos
project\footnote{\texttt{https://nanocosmos.iff.csic.es/}} and installed at the Yebes 40m radio telescope, were used
for the observations of TMC-1. The Q-band receiver consists of two HEMT cold amplifiers covering the
31.0-50.3 GHz band with horizontal and vertical polarisations. Receiver temperatures
vary from 22 K at 32 GHz to 42 K at 50 GHz. The spectrometers are $2\times8\times2.5$ GHz FFTs with a spectral resolution
of 38.1 kHz, providing the whole coverage of the Q-band in both polarisations.
The main beam efficiency and the half power beam width (HPBW) 
of the Yebes 40\,m telescope range from 0.6 and 55$''$ (at 32\,GHz) to 0.43 and 37$''$ (at 49\,GHz), respectively.

The observations leading to the line survey in the Q-band towards TMC-1 
($\alpha_{J2000}=4^{\rm h} 41^{\rm  m} 41.9^{\rm s}$, $\delta_{J2000}=+25^\circ 41' 27.0''$) 
were performed during several sessions
between November 2019 and February 2020. The observing procedure used was
frequency switching with a frequency throw of 10\,MHz. The nominal spectral
resolution of 38.1 kHz was used for the final spectra. 
A study of the
velocity structure of the source \citep[see, e.g.][]{Lique2006,Xue2020} 
could require a higher spectral resolution. However, the
determination of the total column density 
in the line of sight for a given molecule is not affected by our spectral
resolution of 38.1 kHz.
The sensitivity varies along the
Q-band between 1 and 3 mK, which considerably improves previous line surveys in the 31-50 GHz frequency range
\citep{Kaifu2004}.

\begin{figure}[]
\centering
\includegraphics[width=0.8\columnwidth,angle=0]{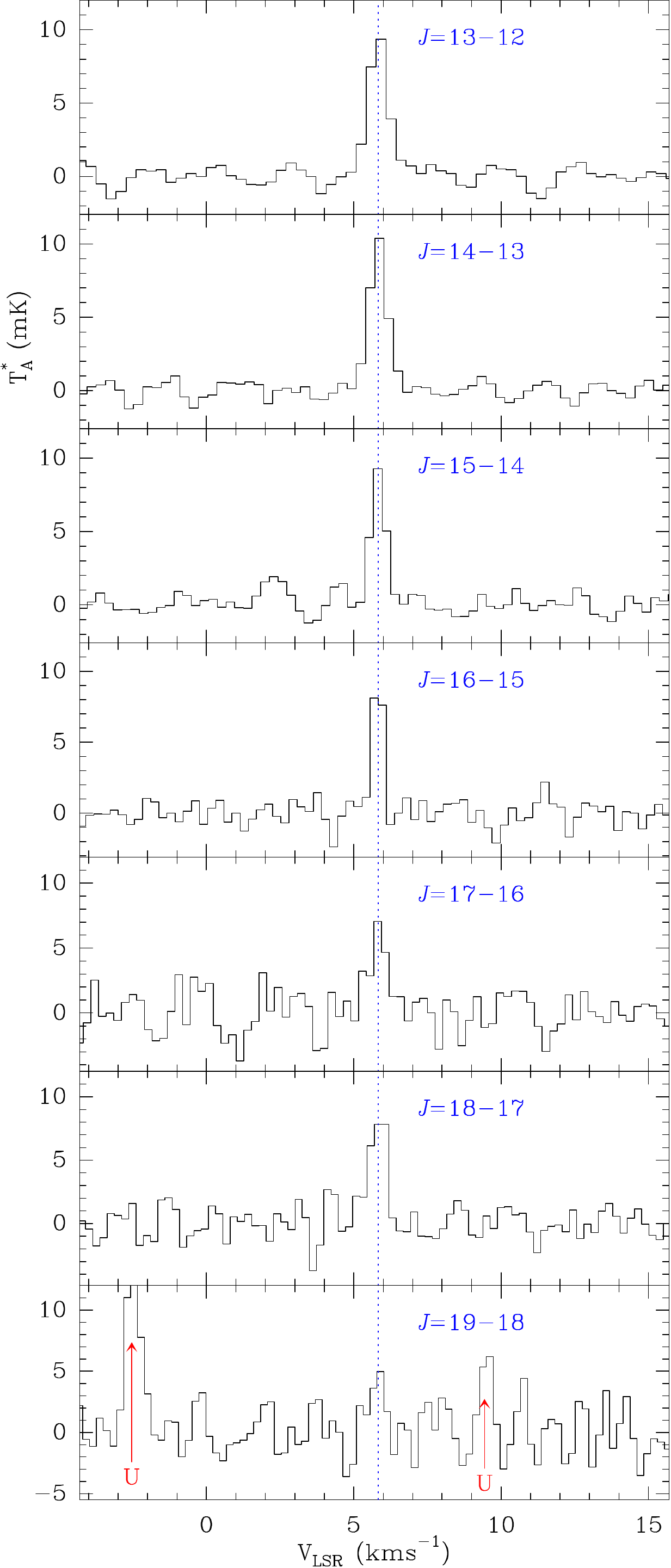}
\caption{Observed lines of the new molecule found in the 31-50 GHz domain towards TMC-1.
The abscissa corresponds to the local standard of rest velocity
in km s$^{-1}$. Frequencies and intensities for the observed lines are given in Table \ref{tab_hc5nhp}.
The ordinate is the antenna temperature corrected for atmospheric and telescope losses in mK. The {\textit J=19-18}
line is only detected at a 3.5$\sigma$ level. Spectral resolution is 38.1 kHz.}
\label{fig_hc5nhp}
\end{figure}

The intensity scale, antenna temperature
($T_A^*$) was calibrated using two absorbers at different temperatures and the
atmospheric transmission model (ATM, \citealt{Cernicharo1985, Pardo2001}).
Calibration uncertainties have been adopted to be 10~\%.
All data have been analysed using the GILDAS package\footnote{\texttt{http://www.iram.fr/IRAMFR/GILDAS}}.

\begin{table}
\small
\caption{Observed line parameters for the new molecule in TMC-1.}
\label{tab_hc5nhp}
\centering
\begin{tabular}{{cccccc}}
\hline \hline
{\textit J$_u$}& $\nu_{obs}^a$& $\nu_{o}-\nu_{c}^b$  & $\Delta$v$^c$ & $\int$T$_A^*$dv $^d$ & T$_A^*$\\
               &  (MHz)       &     (kHz)            & (kms$^{-1}$)& (mK kms$^{-1}$)   & (mK)\\
\hline
13 &33690.975& +5.2 &0.80$\pm$0.05& 8.2$\pm$1.0&  9.7\\
14 &36282.540& -2.6 &0.73$\pm$0.04& 8.0$\pm$1.0& 10.4\\
15 &38874.102& -3.3 &0.59$\pm$0.05& 5.9$\pm$1.0&  9.4\\
16 &41465.655& -3.1 &0.46$\pm$0.05& 4.5$\pm$1.0& 11.9\\
17 &44057.202& +1.7 &0.76$\pm$0.09& 5.1$\pm$1.0&  6.3\\
18 &46648.736& +4.5 &0.68$\pm$0.08& 5.3$\pm$1.0&  7.7\\
19 &49240.242& -8.8 &0.50$\pm$0.10& 2.7$\pm$1.0&  5.0\\
\hline
\end{tabular}
\tablefoot{\\
        \tablefoottext{a}{Observed frequencies for a v$_{LSR}$ of 5.83 km s$^{-1}$. The uncertainty is 10 kHz
    for all lines, except for {\textit J=19-18} for which it is 20 kHz.}\\
        \tablefoottext{b}{Observed minus calculated frequencies in kHz.}\\
        \tablefoottext{c}{Linewidth at half intensity derived by fitting a Gaussian line profile to the observed
     transitions (in kms$^{-1}$).}\\
        \tablefoottext{d}{Integrated line intensity in mK kms$^{-1}$ }\\
}
\end{table}
\normalsize

\section{Results} \label{sec:results}

One of the most surprising results from the line survey in the Q-band in TMC-1 is the presence of
a forest of weak lines. Most of them can be assigned to known species and their isotopologues, and only a few remain unidentified
(Marcelino et al., in preparation). As previously mentioned, the level of sensitivity has been increased by a
factor of 5-10 with respect to previous line surveys performed with other telescopes at these frequencies \citep{Kaifu2004}.
Frequencies for
the unknown lines have been derived by assuming a local standard of rest velocity of 5.83 km s$^{-1}$, which is a value
that was derived from the observed transitions of HC$_5$N and its isotopologues in our line survey.

\subsection{Harmonically related lines}
\label{sec:harmo}
Among the unidentified features in our survey, we have found a series of seven
harmonically related lines to a precision better than 2$\times$10$^{-7}$. This strongly suggests that the carrier is a linear molecule
with a $^1\Sigma$ ground electronic state.
Fig. \ref{fig_hc5nhp} shows these lines and the quantum numbers that were obtained. The derived
line parameters are given in Table \ref{tab_hc5nhp}.
Using the MADEX
code \citep{Cernicharo2012}, we have verified that none of these features can be assigned to
lines from other species.
For a linear molecule, the frequencies of its rotational transitions follow the
standard expression $\nu(J\rightarrow$$J-1)$=2$B_0 J$ - 4$D_0 J^3$.
By fitting the frequencies of the lines  given in Table \ref{tab_hc5nhp},
we derive
\\

$B_0$= 1295.81581(26) MHz

$D_0$=27.4(5)$\times$10$^{-6}$ MHz,
\\
\\
\noindent
where values between parentheses represent the 1$\sigma$ uncertainty on the parameters in units of the
last digit. The standard deviation of the fit is 5.6 kHz.

The rotation constant is $\sim$36 MHz lower than that of HC$_5$N (1331.332 MHz; \citealt{Bizzocchi2004}).
It is also lower than that of dicyanoacetylene, NCCCCN (1336.7 MHz; \citealt{Winther1992}).
The rotational constants of the $^{13}$C and $^{15}$N isotopologues of HC$_5$N are well known
from laboratory measurements \citep{Bizzocchi2004,Giesen2020}. The isotopologues HC$_5^{15}$N and
H$^{13}$CCCCCN have rotational constants of 1298.640 MHz and 1296.676 MHz, respectively,
that is to say they are very close to that of the new species, suggesting that a slightly heavier species could be the
carrier. The isomer HC$_4$NC has a larger rotational
constant of 1401.182 MHz \citep{Botschwina1998} and it can be excluded as a carrier. Moreover, this species
has been detected in our line survey \citep{Cernicharo2020} and also by \citet{Xue2020}.
Although the isomer HNC$_5$ has not been observed in the laboratory, ab initio calculations
indicate that the molecule is bent with a $(B+C)/2$ value of $\sim$1363 MHz \citep{Gronowski2006,Cernicharo2020},
which is $\sim$70 MHz above the observed rotational constant.

In TMC-1, only polyatomic molecules containing H, C, N, O, and S have been found so far. We could consider a linear
chain containing sulphur as a possible carrier. However, the best candidate in this case is
HC$_4$S, which is linear but has a $^2\Pi_i$ ground electronic state and a rotational constant of 1435.326 MHz
that is too high \citep{Hirahara1994}.
Another potential species is \mbox{NCCCS,} which is linear, but also with a $^2\Pi_i$ ground electronic state
and a rotational constant of 1439.186 MHz \citep{McCarthy2003}.
The neutral species HC$_5$O was observed in the laboratory by \citet{Mohamed2005}. It has a rotational constant
of 1293.6 MHz and has been observed in TMC-1 \citep{McGuire2017}. 
Hence, good candidates for the carrier of the observed lines are 
molecular species having a structure and mass close to that of HC$_5$N, HC$_5$O, or NC$_4$N.

The cation HC$_3$NH$^+$ was detected towards TMC-1 by \citet{Kawaguchi1994}; additionally, protonated cyanogen, HNCCN$^+$,
was also detected in this source by \citet{Agundez2015}. Hence, good candidates for the carrier of the series
of harmonically related lines are HC$_5$NH$^+$ and NC$_4$NH$^+$.
An additional candidate is HC$_5$O$^+$, which could be the product of protonation of C$_5$O. However, C$_5$O
has not been detected so far in space and ab initio calculations (see below) indicate a rotational constant
several MHz above the observed one.
For protonated cyanodiacetylene, HC$_5$NH$^+$,
ab initio calculations by \citet{Botschwina1997} indicate an equilibrium rotational constant very close to
ours of $\sim$1294.1 MHz with a dipole moment of 3.811 D. It is a very good candidate indeed as the difference
between the predicted and the observed rotational constant is less than 0.1\%.
Nevertheless, since we have a good second candidate, NC$_4$NH$^+$, 
and also a third (less clear) possibility, HC$_5$O$^+$, 
additional calculations are needed to help with the assignment of the lines.

\subsection{Quantum chemical calculations}

As mentioned before, HC$_5$NH$^+$, NC$_4$NH$^+$, and HC$_5$O$^+$ are the three candidates that have a rotational constant 
compatible with that derived from the lines observed in TMC-1. In order to obtain precise geometries and spectroscopic
molecular parameters that help with the assignment of the observed lines, we carried out high-level 
ab initio calculations for these three species. Other structural isomers of HC$_5$NH$^+$ and NC$_4$NH$^+$ 
species were discarded as candidates because of their rotational constant values and their energetics; all 
metastable isomers lie at least at 18 kcal mol$^{-1}$ over the HC$_5$NH$^+$ and NC$_4$NH$^+$ species, see Table 
\ref{all_isomers}. Therefore, the calculations presented below are restricted to HC$_5$NH$^+$, NC$_4$NH$^+$, and 
HC$_5$O$^+$ molecules. Details regarding the calculations can be found in Appendix \ref{appen_cal}.

The experimental rotational parameters for HC$_5$N, NC$_4$N, and C$_5$O have been determined before 
\citep{Bizzocchi2004,Winther1992,Ogata1995} and, therefore, they can be
used to calibrate the computational results of their corresponding protonated forms. The $B_e$ rotational constant for 
HC$_5$N, NC$_4$N, and C$_5$O, which were calculated at the CCSD(T)-F12/cc-pCVTZ-F12 level of theory, are 1329.57, 1334.69, and 
1363.74 MHz, respectively. The zero-point vibrational contributions to the rotational constant at the MP2/cc-pVTZ 
level of theory were calculated to be 0.62, 0.72, and 1.82 MHz, respectively.  Adding this contribution to the 
above $B_e$, $B_0$ thus takes values of 1330.19, 1335.42, and 1365.16 MHz, respectively, which are very close to the experimental 
values of 1331.332687(20), 1336.68433(30), and 1366.84709(6) MHz. The calculated constants show deviations from 
experimental values by 0.1\% for all the species, which gives confidence as to the accuracy of our calculations.

The $B_0$ rotational constant calculated for HC$_5$NH$^+$, NC$_4$NH$^+$, and HC$_5$O$^+$ were each scaled using the 
ratio $B_{exp}$/$B_{cal}$ for HC$_5$N, NC$_4$N, and C$_5$O, respectively. These values are shown in Table \ref{abini_data} 
together with the estimated values for the centrifugal distortion constant ($D$), which were derived from the frequency calculations 
at CCSD/cc-pVTZ level of theory and scaled in the same manner as the rotational constants. The results of the same calculations
at different levels of theory are given in Tables \ref{abini_datafull} and \ref{abini_test}. Independent of the level 
of theory employed, our calculations provide a rotational constant value for HC$_5$NH$^+$ around 1295.7 MHz, while that 
for NC$_4$NH$^+$ is about 2.0 MHz lower, around 1293.6 MHz. In the case of HC$_5$O$^+$, the value is larger, around 1303.0 
MHz. No large differences were found for the predicted values of the centrifugal distortion constants.

\begin{table}
\tiny
\caption{Rotational constants and electric dipole moments calculated for HC$_5$NH$^+$ and NC$_4$NH$^+$.}
\label{abini_data}
\centering
\begin{tabular}{{|l|l|l|l|l|}}
\hline
\hline
Parameter   & New molecule$^a$ & HC$_5$NH$^+$ & NC$_4$NH$^+$ & HC$_5$O$^+$ \\
\hline
$B_0$ (MHz) & 1295.8158(3)             & 1295.51                &  1293.54             & 1303.12\\
$D_e$ (MHz) & 27.4(5)$\times$10$^{-6}$ &26.3$\times$10$^{-6}$    & 27.8$\times$10$^{-6}$& 31.2$\times$10$^{-6}$     \\
$\mu$ (D) &                          & 3.26                    &   9.47               & 3.10\\
\hline
\end{tabular}
\tablefoot{\\
        \tablefoottext{a}{Values derived from the frequencies observed in TMC-1 (see Section \ref{sec:harmo}). Values between
      parentheses represent the 1~$\sigma$ uncertainty in units of the last digit.}\\
}

\end{table}

\begin{figure}[]
\centering
\includegraphics[width=0.85\columnwidth,angle=0]{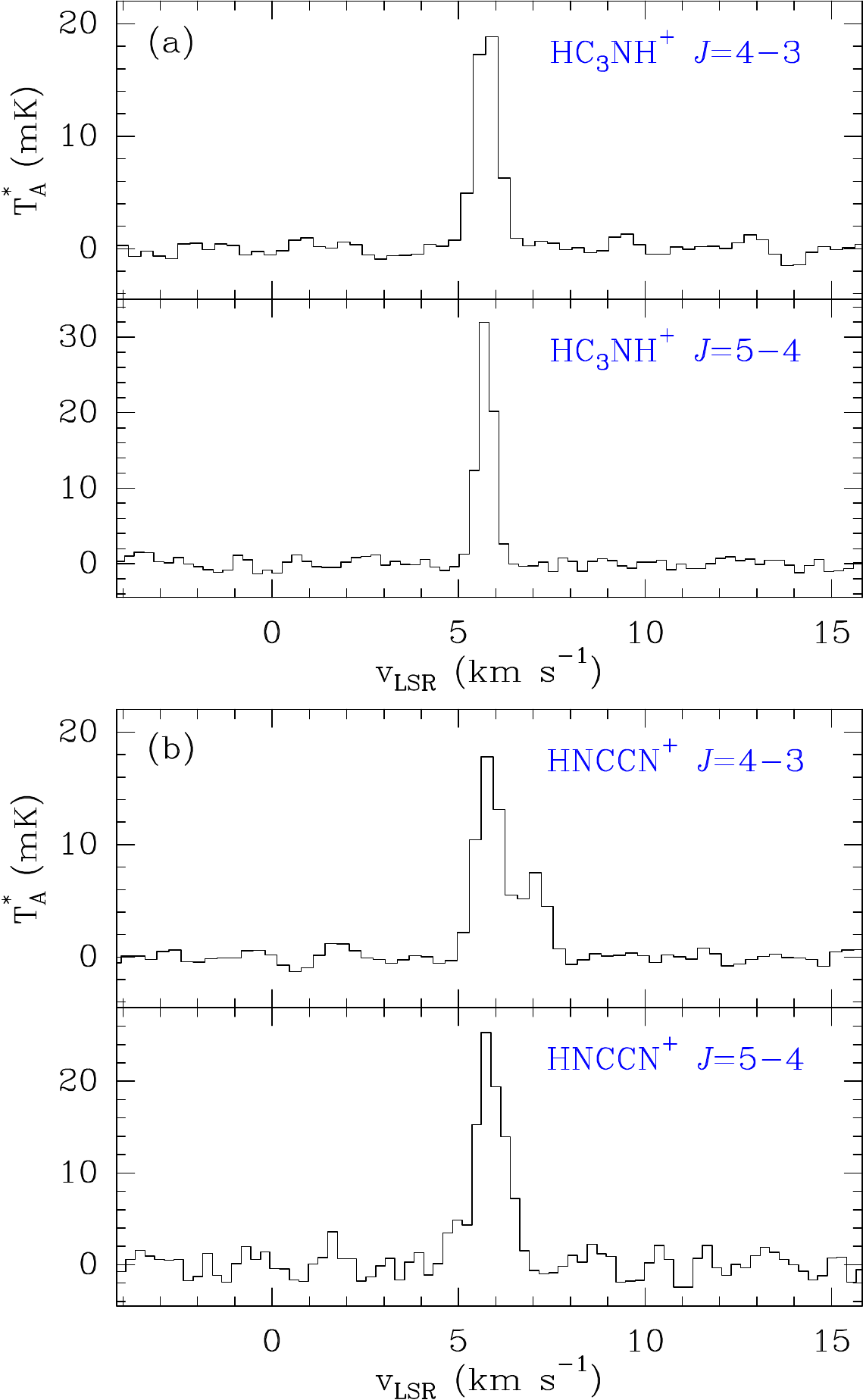}
\caption{Observed lines of HC$_3$NH$^+$ (a) and of HNCCN$^+$ (b).
The abscissa corresponds to the local standard of rest velocity
in km s$^{-1}$. Frequencies and intensities for the observed lines are given in Table \ref{tab_hc3nhp}.
The ordinate is the antenna temperature corrected for atmospheric and telescope losses in mK. The {\textit J=4-3}
line of HNCCN$^+$ shows two components due to the hyperfine structure introduced by the external nitrogen atom.
Spectral resolution is 38.1 kHz.}
\label{fig_hc3nhp}
\end{figure}

\section{Discussion}

The calculated rotational constants of HC$_5$NH$^+$ and NC$_4$NH$^+$ are
very close to the value of $B_0$ observed. A definitive assignment requires observations
in the laboratory. Nevertheless, we could give some support to the assignment to HC$_5$NH$^+$. First,
the ab initio calculations at all levels of theory predict a rotational constant for HC$_5$NH$^+$
around 1295 MHz, while that of NC$_4$NH$^+$ is systematically 2 MHz below. Second, the
rotational constant corrected for the ratio theory-to-observation of the reference species, HC$_5$N
and NC$_4$N, provide an excellent match with the observed $B_0$ for protonated cyanodiacetylene within 0.05\%,
while the difference for protonated dicyanoacetylene reaches 0.2\%.

Additional support for the assignment to HC$_5$NH$^+$ comes
from the comparison of the proton affinities of HC$_5$N and NC$_4$N. As previously indicated,
the abundance of a protonated species depends on the abundance and proton affinity of the neutral counterpart. A large proton affinity permits the transfer of H$^+$ to the
neutral species, M, through the reactions M + XH$^+$ $\rightarrow$ MH$^+$ + X, where
XH$^+$ is an abundant proton donor, such as HCO$^+$ or H$_3^+$.
For cyanodiacetylene, HC$_5$N, the
proton affinity was measured to be 770$\pm$20 kJ mol$^{-1}$ \citep{Edwards2009}. The abundance
of this species is large in cold dark cores, hence, we could expect a moderately high abundance for
its protonated form. Due to the lack of a permanent dipole, species such
as NCCN and NC$_4$N have not been observed in dark clouds so far. Nevertheless, the detection
of NCCNH$^+$ in these objects by \citet{Agundez2015} suggests that NCCN has an abundance as large as $(1-10)\times10^{-8}$ relative to H$_2$, that is to say it is similar to that of HC$_3$N.
Assuming an abundance for NC$_4$N that is similar to that of HC$_5$N, then the relative abundances of
HC$_5$NH$^+$ and NC$_4$NH$^+$  depend on the proton affinities of the neutrals and
on their electronic dissociative recombination rates, which in principle could be assumed to be similar.
For NC$_4$N, there is not an experimental value in the literature, so we calculated it at
CCSD/cc-pVTZ level of theory. We used the
energy balance between NC$_4$N + H$^+$ and NC$_4$NH$^+$, considering NC$_4$N and H$^+$ are independent 
species. We found a proton affinity value for NC$_4$N of 736 kJ/mol.
Using the same scheme, we obtained a proton affinity value for HC$_5$N of 783 kJ/mol,
which is very close to the experimental one (see above). Hence, the proton affinity of NC$_4$N is lower than
that of HC$_5$N, which favours the protonated form of HC$_5$N as a carrier of our lines.

\begin{figure}[]
\centering
\includegraphics[width=0.85\columnwidth,angle=0]{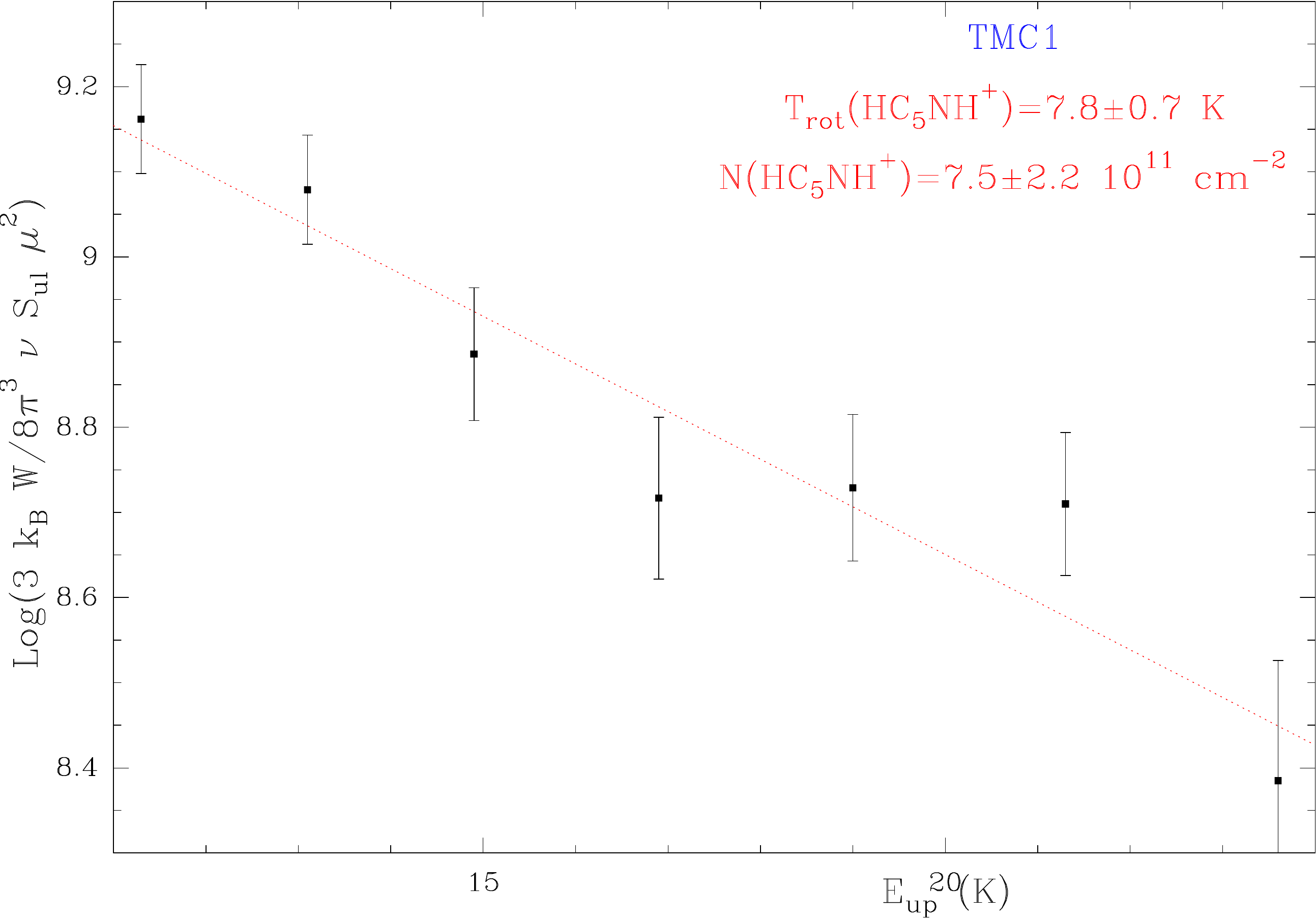}
\caption{Rotational diagram of the observed lines in TMC-1 assuming a dipole
moment of 3.3 D, i.e. assuming the carrier of the lines is HC$_5$NH$^+$.}
\label{fig_hc5nhp_trot}
\end{figure}

Another argument supporting this assignment concerns the
very different dipole moment of the two species. While, HC$_5$NH$^+$ has a predicted dipole moment of
$\sim$3.3 D, the corresponding value for NC$_4$NH$^+$ is $\sim$9.5 D (see Table \ref{abini_data}).
Hence, the abundance resulting from the observed line intensities would be
significantly different if the carrier were to be one versus the other species. In Fig. \ref{fig_hc5nhp_trot} we
show the rotational diagram obtained from the observed intensities (see Table \ref{tab_hc5nhp}) assuming
that the carrier is HC$_5$NH$^+$. We adopted a source radius of 40$''$ \citep{Fosse2001}.
The parameter that is really interesting from this plot is the derived rotational
temperature as the column density depends on the assumed dipole moment. The observed lines can be
reproduced with a {\textit T$_{rot}$} of 7.8 $\pm$ 0.7 K. This is a typical value of the rotational temperature for most
molecules detected so far in TMC-1, but that could require a very large H$_2$ volume density
if the dipole moment were 9.5 D, that is, if the carrier were NC$_4$NH$^+$. The derived column density is (7.5$\pm$2.2)$\times$10$^{11}$ if the
carrier is HC$_5$NH$^+$, and $\sim\,9\times10^{10}$ cm$^{-2}$ if the
carrier is NC$_4$NH$^+$.

Hence, we consider that we
have arguments to support the assignment of the lines to protonated cyanodiacetylene.
\citet{Cernicharo2020} performed a rotational analysis to all the transitions of HC$_5$N
observed in our line survey. They derived {\textit T$_{rot}$}=8.6$\pm$0.2 K and
{\textit N}(HC$_5$N)=(1.8$\pm$0.2)$\times$10$^{14}$ cm$^{-2}$. This column density
was derived from the observed weak hyperfine components in the 
HC$_5$N transitions from $J$=12-11 up to $J$=16-15
in order to take line opacity effects into account. 
Hence, we derived an
abundance ratio of $\sim$240 for HC$_5$N/HC$_5$NH$^+$. 

Another indirect indication supporting our identification
could arise from the comparison of the abundances of other
protonated species, in particular HC$_3$NH$^+$ and NCCNH$^+$. Both species are present
in TMC-1 \citep{Kawaguchi1994,Agundez2015} and two rotational lines of each one are
within our line survey. Fig. \ref{fig_hc3nhp} shows the observed lines of both species. The corresponding
line parameters are given in Table \ref{tab_hc3nhp}.
From the observed intensities, we derived {\textit N}(HC$_3$NH$^+$)=1.0$\times$10$^{12}$ cm$^{-2}$
and {\textit N}(HNCCN$^+$)=9.0$\times$10$^{10}$ cm$^{-2}$, that is, {\textit N}(HC$_3$NH$^+$)/{\textit N}(HNCCN$^+$)$\sim$14.
\citet{Cernicharo2020} also derived a column density for HC$_3$N of 
(2.3$\pm$0.2)$\times$10$^{14}$ cm$^{-2}$, which was corrected for line opacity effects. 
Therefore, the
derived abundance ratio for HC$_3$N/HC$_3$NH$^+$ is $\simeq$230, 
which is a value that is in good agreement with the one found by
\citet{Kawaguchi1994} $\sim$160 and it is very similar to what was found above 
for HC$_5$N/HC$_5$NH$^+$.

Between our unidentified features, we searched  for lines that could be assigned to NC$_4$NH$^+$ using our ab initio
calculations without success.
A search for lines of NCCCNC also provides an upper limit to its column density of $\le$10$^{12}$ cm$^{-2}$.

The case for HC$_5$NH$^+$ is similar to that of the discovery of C$_3$H$^+$ by \citet{Pety2012},
which was done from the reasonable agreement between the observed rotational constant and the
best  ab initio calculations available at that moment for C$_3$H$^+$.

\begin{table}
\tiny
\caption{Observed line parameters for HC$_3$NH$^+$ and HNCCN$^+$.}
\label{tab_hc3nhp}
\centering
\begin{tabular}{{cccccc}}
\hline \hline
{\textit J$_u$}& $\nu_{rest}^a$&  v$_{LSR}$        & $\Delta$v & $\int$T$_A^*$dv & T$_A^*$\\
               &  (MHz)        &     (km\,s$^{-1}$) & (km\,s$^{-1}$)& (mK km\,s$^{-1}$)  & (mK)\\
\hline
HC$_3$NH$^+$\\
4      &34631.859& 5.76$\pm$0.02&   0.72$\pm$0.02& 16.1$\pm$1&  20.9\\ 
5      &43289.740& 5.74$\pm$0.02&   0.52$\pm$0.01& 18.1$\pm$1&  32.7\\ 
\hline
HNCCN$^+$\\
4$^b$  &35503.840& 5.86$\pm$0.03&   0.76$\pm$0.07&  6.1$\pm$1&  7.5\\
4$^c$  &35503.981& 5.82$\pm$0.01&   0.82$\pm$0.03& 15.7$\pm$1& 18.0\\ 
5      &44379.851& 5.86$\pm$0.02&   0.90$\pm$0.04& 22.9$\pm$1& 23.8\\ 
\hline
\end{tabular}
\tablefoot{\\
        \tablefoottext{a}{Rest frequencies. The uncertainty is typically 2-3 kHz.}\\
        \tablefoottext{b}{Corresponds to the {\textit F=3-2} hyperfine component.}\\
        \tablefoottext{c}{Corresponds to the unresolved hyperfine components
    {\textit F=4-3 \& 5-4}.}\\
}
\normalsize
\end{table}

\subsection{Chemistry of protonated species}

\begin{figure}[]
\centering
\includegraphics[width=\columnwidth,angle=0]{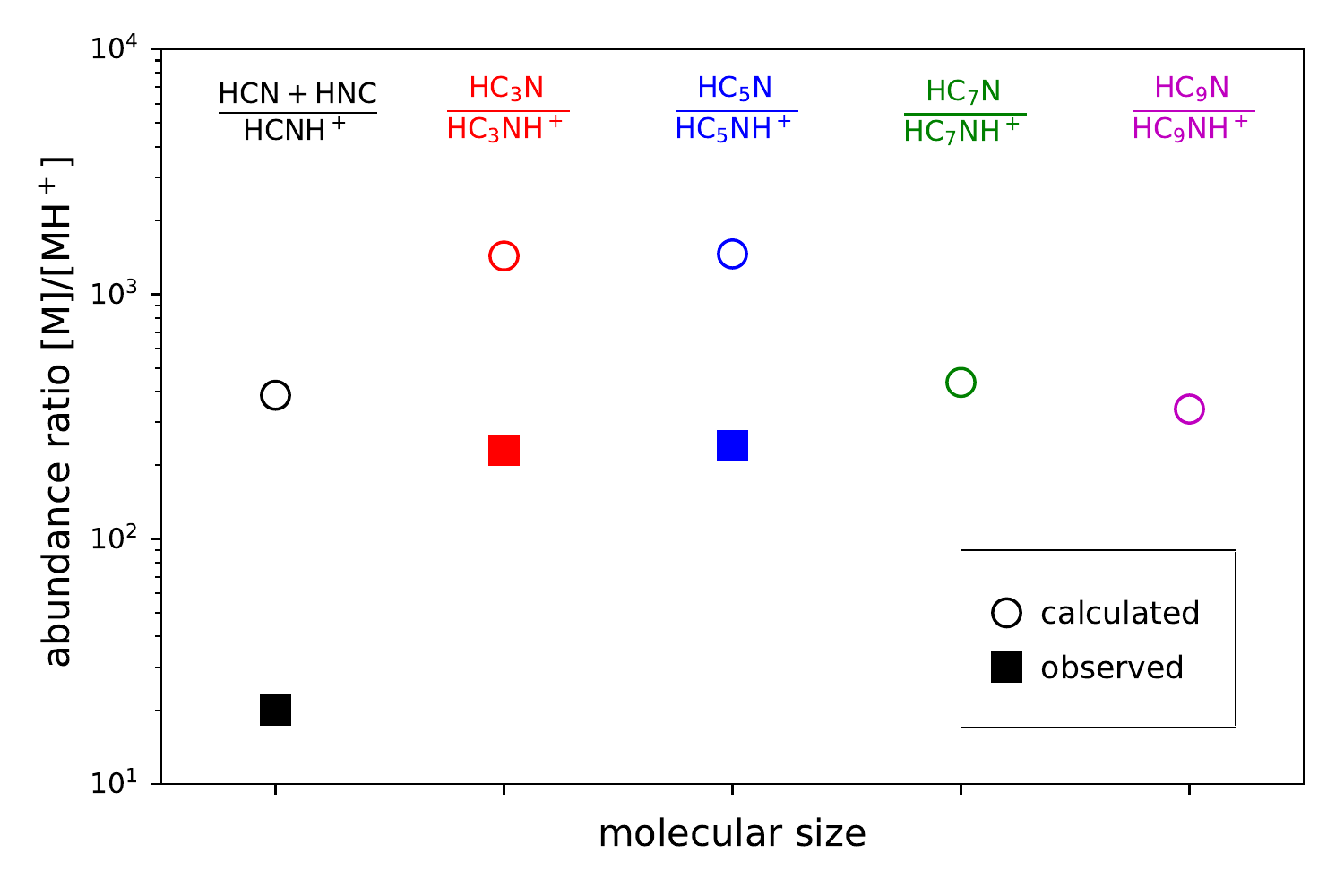}
\caption{Abundance ratios between neutral and protonated species for the series of cyanopolyynes.
Calculated values correspond to a chemical model of a cold dark cloud at the steady state and observed ones to TMC-1.}
\label{fig_ratios}
\end{figure}

The chemistry of protonated molecules in cold dense clouds has been discussed by \cite{Agundez2015}. Here, we revisit the pseudo time-dependent gas-phase chemical model of a cold dark cloud carried out by these authors. The chemical network adopted is largely based on the UMIST {\small RATE12} reaction network \citep{McElroy2013}. For more details on the chemical model, we refer the reader to \cite{Agundez2015}. In cold dense clouds, protonated molecules MH$^+$ form mainly by the proton transfer from a proton donor XH$^+$ to M :
\begin{equation}
\rm XH^+ + M \rightarrow MH^+ + X, \label{reac:proton-transfer}
\end{equation}
while they are destroyed through the dissociative recombination with electrons
\begin{equation}
\rm MH^+ + e^- \rightarrow products. \label{reac:dissociative-recombination}
\end{equation}
Therefore, at the steady state, the neutral-to-protonated abundance ratio is
\begin{equation}
{\rm \frac{[M]}{[MH^+]}} = \frac{k_{DR}}{k_{PT}} {\rm \frac{[e^-]}{[XH^+]}}, \label{eq:abundance-ratio}
\end{equation}
where $k_{DR}$ and $k_{PT}$ are the rate constants of the reactions of the dissociative recombination and proton transfer,
respectively. The chemical model indicates that this scheme holds for HC$_3$NH$^+$ and HC$_5$NH$^+$, where the
main proton donor XH$^+$ is HCO$^+$. Calculated neutral-to-protonated abundance ratios are about ten times higher
than observed for HC$_3$NH$^+$ and HC$_5$NH$^+$ (see Fig.~\ref{fig_ratios}). That is to say, the chemical model underestimates
the abundance of the protonated form with respect to its neutral counterpart. The rate constants of the dissociative
recombination and proton transfer from HCO$^+$ used in the chemical model are just estimates for HC$_5$NH$^+$;
although, in the case of HC$_3$NH$^+$, they are well known from experiments \citep{Anicich2003,Geppert2004}. 
We note
that the chemical model also underestimates the abundance of HCNH$^+$ (see Fig.~\ref{fig_ratios}) and other protonated
species, such as HCS$^+$ and NH$_3$, regardless of the cosmic-ray ionisation adopted \citep{Agundez2015}. We suspect that
the most likely reason for the underestimation of protonated molecules in the chemical model is the existence of additional
formation routes that are not considered in the model.

\begin{acknowledgements}

The Spanish authors thank Ministerio de Ciencia e Innovaci\'on for funding
support through project AYA2016-75066-C2-1-P. We also thank ERC for funding through grant
ERC-2013-Syg-610256-NANOCOSMOS. MA thanks Ministerio de Ciencia e Innovaci\'on
for Ram\'on y Cajal grant RyC-2014-16277.
\end{acknowledgements}

\begin{appendix}
\section{Frequencies, intensities, and energies of HC$_5$NH$^+$ transitions}
\label{appen_freq}

In order to facilitate the search of HC$_5$NH$^+$ towards other sources, 
we provide
the frequencies, uncertainties, line strenghts, Einstein coefficients, and upper energy
levels for all transitions of this species  below 100 GHz in Table \ref{freq_hc5nh+}. The denegeracies
of the energy levels are 2$J$+1. The rotational partition function for HC$_5$NH$^+$ 
can be derived for a given temperature T from the standard expresion 
Q$_{rot}$=kT/hB \citep{Gordy1984}, to be Q$_{rot}$=16.087$\times$T.

\begin{table}
\caption{Frequencies, upper energy levels, Einstein coefficients, and line strengths of HC$_5$NH$^+$ transitions.}
\label{freq_hc5nh+}
\centering
\small
\begin{tabular}{{|c|c|c|c|c|}}
\hline
 $J_u-J_l$ &  $\nu$ (MHz)      &E$_{up}$(K)& $A_{ul}$ (s$^{-1}$)&$S_{ul}$\\
\hline                          
  1- 0   &   2591.632$\pm$0.001&  0.1    & 7.177$\times$10$^{-10}$&    1 \\
  2- 1   &   5183.262$\pm$0.001&  0.4    & 6.890$\times$10$^{-09}$&    2 \\
  3- 2   &   7774.892$\pm$0.002&  0.7    & 2.492$\times$10$^{-08}$&    3 \\
  4- 3   &  10366.520$\pm$0.002&  1.2    & 6.125$\times$10$^{-08}$&    4 \\
  5- 4   &  12958.144$\pm$0.002&  1.9    & 1.223$\times$10$^{-07}$&    5 \\
  6- 5   &  15549.766$\pm$0.003&  2.6    & 2.147$\times$10$^{-07}$&    6 \\
  7- 6   &  18141.384$\pm$0.003&  3.5    & 3.447$\times$10$^{-07}$&    7 \\
  8- 7   &  20732.997$\pm$0.003&  4.5    & 5.188$\times$10$^{-07}$&    8 \\
  9- 8   &  23324.605$\pm$0.003&  5.6    & 7.435$\times$10$^{-07}$&    9 \\
 10- 9   &  25916.207$\pm$0.003&  6.8    & 1.025$\times$10$^{-06}$&   10 \\
 11-10   &  28507.802$\pm$0.003&  8.2    & 1.371$\times$10$^{-06}$&   11 \\
 12-11   &  31099.391$\pm$0.003&  9.7    & 1.786$\times$10$^{-06}$&   12 \\
 13-12   &  33690.971$\pm$0.003& 11.3    & 2.278$\times$10$^{-06}$&   13 \\
 14-13   &  36282.543$\pm$0.003& 13.1    & 2.852$\times$10$^{-06}$&   14 \\
 15-14   &  38874.105$\pm$0.003& 14.9    & 3.516$\times$10$^{-06}$&   15 \\
 16-15   &  41465.658$\pm$0.003& 16.9    & 4.276$\times$10$^{-06}$&   16 \\
 17-16   &  44057.200$\pm$0.003& 19.0    & 5.138$\times$10$^{-06}$&   17 \\
 18-17   &  46648.732$\pm$0.003& 21.3    & 6.109$\times$10$^{-06}$&   18 \\
 19-18   &  49240.251$\pm$0.004& 23.6    & 7.195$\times$10$^{-06}$&   19 \\
 20-19   &  51831.758$\pm$0.006& 26.1    & 8.402$\times$10$^{-06}$&   20 \\
 21-20   &  54423.251$\pm$0.008& 28.7    & 9.738$\times$10$^{-06}$&   21 \\
 22-21   &  57014.731$\pm$0.010& 31.5    & 1.121$\times$10$^{-05}$&   22 \\
 23-22   &  59606.197$\pm$0.012& 34.3    & 1.282$\times$10$^{-05}$&   23 \\
 24-23   &  62197.647$\pm$0.015& 37.3    & 1.458$\times$10$^{-05}$&   24 \\
 25-24   &  64789.082$\pm$0.018& 40.4    & 1.649$\times$10$^{-05}$&   25 \\
 26-25   &  67380.500$\pm$0.021& 43.7    & 1.856$\times$10$^{-05}$&   26 \\
 27-26   &  69971.902$\pm$0.025& 47.0    & 2.080$\times$10$^{-05}$&   27 \\
 28-27   &  72563.285$\pm$0.029& 50.5    & 2.322$\times$10$^{-05}$&   28 \\
 29-28   &  75154.650$\pm$0.033& 54.1    & 2.581$\times$10$^{-05}$&   29 \\
 30-29   &  77745.996$\pm$0.038& 57.8    & 2.859$\times$10$^{-05}$&   30 \\
 31-30   &  80337.323$\pm$0.043& 61.7    & 3.156$\times$10$^{-05}$&   31 \\
 32-31   &  82928.629$\pm$0.048& 65.7    & 3.473$\times$10$^{-05}$&   32 \\
 33-32   &  85519.914$\pm$0.054& 69.8    & 3.811$\times$10$^{-05}$&   33 \\
 34-33   &  88111.177$\pm$0.060& 74.0    & 4.170$\times$10$^{-05}$&   34 \\
 35-34   &  90702.419$\pm$0.067& 78.4    & 4.550$\times$10$^{-05}$&   35 \\
 36-35   &  93293.637$\pm$0.073& 82.8    & 4.953$\times$10$^{-05}$&   36 \\
 37-36   &  95884.831$\pm$0.081& 87.4    & 5.380$\times$10$^{-05}$&   37 \\
 38-37   &  98476.002$\pm$0.089& 92.2    & 5.830$\times$10$^{-05}$&   38 \\
\hline
\end{tabular}
\end{table}
\normalsize

\section{Additional quantum chemical calculation data}
\label{appen_cal}

All structure optimisation calculations reported in this work were performed using the closed-shell
coupled cluster with singles and doubles (CCSD) \citep{Cizek1969} and perturbative triple
excitations (CCSD(T), \citealt{Raghavachari1989}) with and without an explicitly correlated (F12)
approximation \citep{Adler2007,Knizia2009}. For CCSD and CCSD(T) calculations, we used Dunning’s
correlation consistent polarised valence (and valence-core) triple-$\zeta$ basis sets cc-pVTZ
(cc-pCVTZ), which were also augmented with diffuse functions (aug-cc-pVTZ,  \citealt{Pritchard2019}). On the other hand, with the calculations at the CCSD(T)-F12 level, the Dunning's correlation consistent basis sets with polarised core-valence correlation triple-$\zeta$ for explicitly correlated calculations (cc-pCVTZ-F12; \citealt{Hill2010a,Hill2010b}) was used. In this latter case, all electrons (valence and core) are correlated. In order to achieve an estimate of the $B_0$ rotational constant, vibration-rotation interaction constants were calculated using second-order perturbation
theory at the MP2/cc-pVTZ level. All
the calculations were carried out using the Molpro 2018.1 \citep{Werner2018} and Gaussian 09 \citep{Frisch2013}
programme packages.

The molecular structure for all the linear isomers of HC$_5$NH$^+$ and NC$_4$NH$^+$ species were
calculated using CCSD/cc-pVTZ level of theory. We did not considered asymmetric structures because
the carrier of the observed lines is a linear molecule. The relative energies of all plausible isomers,
as well as their rotational constants and dipole moments obtained at the CCSD/cc-pVTZ level, are summarised
in Table \ref{all_isomers}. In both cases (C$_5$H$_{2}$N$^+$ and C$_4$HN$_{2}^+$), the considered species
HC$_5$NH$^+$ and NC$_4$NH$^+$ are the lowest energy isomers, respectively.

\begin{table*}
\small
\caption{Theoretical values for spectroscopic parameters of HC$_5$NH$^+$, NC$_4$NH$^+$, and HC$_5$O$^+$ at different levels of theory.}
\label{abini_datafull}
\centering
\begin{tabular}{{lccccccccc}}
\hline
\hline
&\multicolumn{3}{c}{HC$_5$NH$^+$}&\multicolumn{3}{c}{NC$_4$NH$^+$} &\multicolumn{3}{c}{HC$_5$O$^+$} \\
\cmidrule(lr){2-4} \cmidrule(lr){5-7} \cmidrule(lr){8-10}
& $B_0$\tablefootmark{a} (MHz)  & $D$\tablefootmark{b} (MHz) & $\mu$ (D) &  $B_0$\tablefootmark{c} (MHz)  & $D$\tablefootmark{d} (MHz) & $\mu$ (D) &  $B_0$\tablefootmark{e} (MHz)  & $D$\tablefootmark{f} (MHz) & $\mu$ (D)\\
\hline
CCSD(T)-F12;core/CVTZ-F12   & 1295.51  &                         &  3.26 &  1293.54 &                        & 9.47  &  1303.1   &                          &    3.13   \\
CCSD(T)/aug-cc-pVTZ         & 1295.98  &                         &  3.22 &  1293.95 &                        & 9.46  &  1303.4   &                          &    3.22   \\
CCSD(T)/cc-pCVTZ            & 1295.78  &                         &  3.83 &  1293.77 &                        & 9.47  &  1303.1   &                          &    2.22   \\
CCSD(T)/cc-pVTZ             & 1295.77  &                         &  3.82 &  1293.87 &                        & 9.45  &  1309.1   &                          &    2.23   \\
CCSD/aug-cc-pVTZ            & 1295.93  &                         &  3.82 &  1293.54 &                        & 9.94  &  1297.9   &                          &    2.78   \\
CCSD/cc-pCVTZ               & 1295.73  &                         &  3.53 &  1293.46 &                        & 9.93  &  1298.5   &                          &    1.63   \\
CCSD/cc-pVTZ                & 1295.67  &   26.3$\times$10$^{-6}$ &  3.55 &  1293.50 & 27.8$\times$10$^{-6}$  & 9.91  &  1297.9   & 31.2$\times$10$^{-6}$    &    1.65   \\
MP2/cc-pVTZ                 & 1296.20  &   26.1$\times$10$^{-6}$ &  3.64 &  1293.41 & 27.6$\times$10$^{-6}$  & 9.13  &  1304.7   & 30.7$\times$10$^{-6}$    &    2.07   \\

\hline
\end{tabular}
\tablefoot{
\tablefoottext{a}{Rotational constants were corrected with the vibration-rotation interaction estimated at the MP2/cc-pVTZ level of theory and by the scaling factors found for HC$_5$N.} \tablefoottext{b}{Centrifugal distortion constant values were corrected using the scaling factor found for those calculated for HC$_5$N.} \tablefoottext{c}{Rotational constants were corrected with the vibration-rotation interaction estimated at the MP2/cc-pVTZ level of theory and by the scaling factors found for NC$_4$N.} \tablefoottext{d}{Centrifugal distortion constant values were corrected using the scaling factor found for those values calculated for NC$_4$N.}
\tablefoottext{e}{Rotational constants were corrected with the vibration-rotation interaction estimated at the MP2/cc-pVTZ level of theory and by the scaling factors found for C$_5$O.} \tablefoottext{f}{Centrifugal distortion constant values were corrected using the scaling factor found for those values calculated for C$_5$O.}  }
\end{table*}

\begin{table*}
\caption{Comparison of the experimental and theoretical values for spectroscopic parameters of HC$_5$N, NC$_4$N, and C$_5$O at different levels of theory.}
\label{abini_test}
\centering
\begin{tabular}{{lccc}}
\hline
\hline
&\multicolumn{3}{c}{HC$_5$N}\\
\cmidrule(lr){2-4}
& $B_0$\tablefootmark{a} (MHz)  & $D$\tablefootmark{b} (MHz) & $\mu$ (D)\\
Experimental                & 1331.332687(20)  &   30.1090 (15)$\times$10$^{-6}$ &   4.33\tablefootmark{c} \\
\hline
CCSD(T)-F12;core/CVTZ-F12   &  1330.19         &                                 &   4.88                  \\
CCSD(T)/aug-cc-pVTZ         &  1317.27         &                                 &   4.90                  \\
CCSD(T)/cc-pCVTZ            &  1318.29         &                                 &   4.33                  \\
CCSD(T)/cc-pVTZ             &  1317.73         &                                 &   4.33                  \\
CCSD/aug-cc-pVTZ            &  1324.51         &                                 &   4.84                  \\
CCSD/cc-pCVTZ               &  1325.30         &                                 &   4.32                  \\
CCSD/cc-pVTZ                &  1324.75         &   27.0$\times$10$^{-6}$         &   4.32                  \\
MP2/cc-pVTZ                 &  1320.44         &   26.8$\times$10$^{-6}$         &   4.35                  \\
\hline
\hline
&\multicolumn{2}{c}{NC$_4$N\tablefootmark{d}}    \\
\cmidrule(lr){2-4}
& $B_0$\tablefootmark{a} (MHz)  & $D$\tablefootmark{b} (MHz)\\
Experimental              & 1336.68433(30)  &  31.44 (12) $\times$10$^{-6}$ &  \\
\hline
CCSD(T)-F12;core/CVTZ-F12 &    1330.13      &                               &  \\
CCSD(T)/aug-cc-pVTZ       &    1322.36      &                               &  \\
CCSD(T)/cc-pCVTZ          &    1323.40      &                               &  \\
CCSD(T)/cc-pVTZ           &    1322.90      &                               &  \\
CCSD/aug-cc-pVTZ          &    1330.13      &                               &  \\
CCSD/cc-pCVTZ             &    1330.86      &                               &  \\
CCSD/cc-pVTZ              &    1330.46      & 28.3$\times$10$^{-6}$         &  \\
MP2/cc-pVTZ               &    1323.60      & 28.1$\times$10$^{-6}$         &  \\
\hline
\hline

&\multicolumn{3}{c}{C$_5$O}                       \\
\cmidrule(lr){2-4}
& $B_0$\tablefootmark{a} (MHz)  & $D$\tablefootmark{b} (MHz) & $\mu$ (D) \\

Experimental             &    1366.84709(6)    &   35.05 (51)$\times$10$^{-6}$      & \\ 
\hline
CCSD(T)-F12;core/CVTZ-F12&        1365.56           &                                   &  3.04    \\
CCSD(T)/aug-cc-pVTZ      &        1350.44           &                                   &  2.92    \\
CCSD(T)/cc-pCVTZ         &        1351.38           &                                   &  4.01    \\
CCSD(T)/cc-pVTZ          &        1349.51           &                                   &  3.99    \\
CCSD/aug-cc-pVTZ         &        1365.85           &                                   &   4.34   \\
CCSD/cc-pCVTZ            &        1366.78           &                                   &   4.38   \\
CCSD/cc-pVTZ             &        1365.27           &   26.4$\times$10$^{-6}$           &   4.36   \\
MP2/cc-pVTZ              &        1350.69           &   25.9$\times$10$^{-6}$           &   4.49   \\

\hline
\end{tabular}
\tablefoot{
\tablefoottext{a}{Rotational constants were corrected with the vibration-rotation interaction estimated at the MP2/cc-pVTZ level of theory (see text of section \ref{appen_cal}).} \tablefoottext{b}{Centrifugal distortion constant was only calculated at  CCSD/cc-pVTZ and  MP2/cc-pVTZ levels of theory.} \tablefoottext{c}{\cite{Alexander1976}.}
\tablefoottext{d}{NC$_{4}$N is a non-polar species.} }
\end{table*}

\begin{table}
\small
\caption{Relative energies, rotational constants, and electric dipole moments for HC$_5$NH$^+$ and NC$_4$NH$^+$ and all their isomers calculated at the CCSD/cc-pVTZ level of theory.}
\label{all_isomers}
\centering
\begin{tabular}{{l|ccc}}
        \hline\hline
        Isomer       & $\Delta$$E$ (kcal/mol) & $B_e$ (MHz) & $\mu$ (D) \\ \hline
        C$_5$H$_{2}$N$^+$ &  \\ \hline
        HCCCCCNH$^+$ & 0                      & 1288.24     & 4.33      \\
        HCCCNCCH$^+$ & 18.47                  & 1385.97     & 1.22      \\
        HCCCCNCH$^+$ & 20.99                  & 1342.20     & 5.98      \\ \hline\hline
        C$_4$HN$_{2}^+$ &  \\ \hline
        HNCCCCN$^+$  & 0                      & 1286.43     & 9.91      \\
        HNCCCNC$^+$  & 18.44                  & 1363.81     & 7.87      \\
        HCCCNCN$^+$  & 21.03                  & 1392.43     & 6.50      \\
        HCCNCCN$^+$  & 25.67                  & 1387.78     & 4.36      \\
        HCCCCNN$^+$  & 37.08                  & 1357.83     & 0.15      \\
        HNCNCCC$^+$  & 64.66                  & 1423.10     & 6.90      \\
        HCCCNNC$^+$  & 64.76                  & 1457.42     & 4.72      \\
        HNCCNCC$^+$  & 94.42                  & 1408.11     & 8.86      \\
        HCNNCCC$^+$  & 115.32                 & 1479.97     & 9.38      \\
        HNNCCCC$^+$  & 140.49                 & 1382.25     & 6.48      \\
        HCCNNCC$^+$  & 145.28                 & 1512.05     & 4.01      \\ \hline
\end{tabular}
\end{table}

Vibrational calculations were also conducted in order to estimate the energies, the
IR intensities, and the first order vibration-rotation coupling constant (Table \ref{Vibrations}). From these values, we derived the rotational constants
$B_{\nu}$ for each vibrational state following the expression \citep{Gordy1984}:
\begin{equation} \label{eq_Bv1}
B_{\nu}= B_{e} - \sum_i \alpha_i (v_i + \tfrac{1}{2}d_{i})
,\end{equation}
where $B_{e}$ is the rotational constant at the equilibrium position and $\alpha_{i}$ represents
the first order vibration-rotation coupling constants for each $i$ vibrational mode. Keeping
in mind that the rotational constant of the vibrational ground state ($B_{0}$) is obtained when
the $v$ for every vibrational mode is equal to zero ($B_{0}= B_{e} - \sum_i
\tfrac{1}{2}\alpha_id_{i}$), equation \ref{eq_Bv1} can be also expressed as
$B_{\nu}= B_{0} - \sum_i \alpha_i v_i$.

The values for the rotational constants of the fundamental vibrations, $B_{\nu}$, of
HC$_{5}$NH$^{+}$ and HNC$_{4}$N$^{+}$ were corrected by the scale factor for HC$_{5}$N
and NC$_{4}$N, respectively, which were obtained as follows: $B_{0}$, the experimental value (\citealp{Bizzocchi2004,
Winther1992}, respectively), was divided by the theoretical values (MP2/cc-pVTZ level of theory for $B_{\nu}$
estimations). In the case of the $B_{0}$ values reported in Table \ref{abini_test}, we initially calculated
the theoretical ground state rotational constant $B_{0}$ ($B_{0}= B_{e} - \sum_i  \tfrac{1}{2}\alpha_id_{i}$),
using the $B_{e}$ value of the corresponding level of theory and the vibration-rotation coupling constants
($\alpha$) from the MP2/cc-pVTZ anharmonic calculations, and, finally, we corrected the $B_{0}$ values
using a scale factor of HC$_{5}$N and NC$_{4}$N at  the corresponding ab~initio level of calculation.

From the energies of the vibrational modes, which were obtained theoretically, we could also estimate the vibrational
partition function for the HC$_{5}$NH$^{+}$ and HNC$_{4}$N$^{+}$ species. They were calculated using the
expression \citep{Gordy1984}:
\begin{equation}
Q_{v} = \prod_{i}(1-e^{-(h\omega_{i}/KT)})^{-d_{i}}
,\end{equation}
where $\omega_{i}$ and $d_{i}$ represent the energy and the degeneracy of each $i$ vibrational mode.
These energies for the vibrational modes are taken from the results of the ab initio calculations evaluated
at the MP2/cc-pVTZ level of theory, under the anharmonic correction.

\begin{table*}  
        \caption{Rotational constants, energies, and IR Intensities for the vibrational modes of HC$_{5}$NH$^{+}$
    and HNC$_{4}$N$^{+}$.}\label{Vibrations}
        \centering
        \begin{tabular}{{ccrccccrccc}}
                \hline  
                \hline
                &\multicolumn{5}{c}{HC$_{5}$NH$^{+}$ \tablefootmark{(a)}} &\multicolumn{5}{c}{HNC$_{4}$N$^{+}$ \tablefootmark{(a)}} \\
                \cmidrule(lr){2-6} \cmidrule(lr){7-11}
                Mode & E$_{\nu}$ \tablefootmark{(b)} & I$_{IR} $\tablefootmark{(c)} & $\alpha_{\nu}$ \tablefootmark{(d)} & B$_{\nu}$ \tablefootmark{(e)}& q$^{e}_{\nu}$ \tablefootmark{(f)}& E$_{\nu}$ \tablefootmark{(b)} & I$_{IR}$ \tablefootmark{(c)}&  $\alpha_{\nu}$ \tablefootmark{(d)} &B$_{\nu}$ \tablefootmark{(e)}& q$^{e}_{\nu}$ \tablefootmark{(f)} \\ \hline
                $\nu_{1}$    & 3694 & 1382.293 & ~1.048 & 1295.142 &         & 3491 & 1260.325 & ~1.044 & 1293.961 &         \\
                $\nu_{2}$    & 3428 & 140.639  & ~0.881 & 1295.310 &         & 2240 & 513.091  & ~3.812 & 1291.164 &         \\
                $\nu_{3}$    & 2295 & 1057.223 & ~3.898 & 1292.267 &         & 2130 & 363.857  & ~4.319 & 1290.651 &         \\
                $\nu_{4}$    & 2182 & 659.964  & ~4.183 & 1291.981 &         & 1992 & 52.358   & ~2.924 & 1292.061 &         \\
                $\nu_{5}$    & 2021 & 60.844   & ~2.546 & 1293.631 &         & 1194 & 26.120   & ~2.879 & 1292.107 &         \\
                $\nu_{6}$    & 1236 & 7.209    & ~2.597 & 1293.580 &         & 608  & 23.291   & ~1.041 & 1293.964 &         \\
                $\nu_{7}$    & 630  & 30.504   & ~0.963 & 1295.228 &         & 523  & 63.591   & -1.082 & 1296.109 & 0.2652  \\
                $\nu_{8}$    & 711  & 32.967   & -0.160 & 1296.361 & 0.1826  & 502  & 28.043   & -1.414 & 1296.444 & 0.2750  \\
                $\nu_{9}$    & 529  & 15.163   & -1.594 & 1297.807 & 0.2725  & 502  & 44.205   & -1.103 & 1296.130 & 0.2652  \\
                $\nu_{10}$   & 509  & 9.260    & -1.295 & 1297.505 & 0.2786  & 241  & 10.300   & -2.579 & 1297.622 & 0.4563  \\
                $\nu_{11}$   & 408  & 122.178  & -0.796 & 1297.002 & 0.2859  & 96   & 0.431    & -2.888 & 1297.934 & 1.1186  \\
                $\nu_{12}$   & 254  & 0.168    & -2.366 & 1298.585 & 0.4631  &      &          &        &          &         \\
                $\nu_{13}$   & 99   & 11.423   & -2.877 & 1299.101 & 1.1384  &      &          &        &          &         \\  \hline
        \end{tabular}           
        \tablefoot{\\
                \tablefoottext{a}{Results of ab initio calculations evaluated under anharmonic correction at the MP2/cc-pVTZ level of theory.} \\
                \tablefoottext{b}{Energy of the corresponding ${\nu}$ vibrational mode in cm$^{-1}$.}\\
                \tablefoottext{c}{Infrared intensity in units of km mol$^{-1}$.}\\
                \tablefoottext{d}{Vibration-rotation coupling constants $\alpha$ estimated for each $\nu$  vibrational mode.}\\
                \tablefoottext{e}{Rotational constant $B$ of the vibrational state $\nu$ estimated using the $\alpha$ values and corrected with the corresponding experimental B$_{0}$ (see text).}\\
                \tablefoottext{f}{$l$-doubling constant $q^{e}_{\nu}$ of the vibrational mode $\nu$.}\\
        }

\end{table*}

\begin{table}
\caption{Vibrational partition function for HC$_{5}$NH$^{+}$  and HNC$_{4}$N$^{+}$ species. } \label{VibParFunMP2}
\centering
\begin{tabular}{{ccc}}
\hline\hline
        Temperature (K) & HC$_{5}$NH$^{+}$  & HNC$_{4}$N$^{+}$  \\ \hline
        300        &     30.61         &      28.61        \\
        100        &     1.85          &      1.91         \\
        80         &     1.48          &      1.52         \\
        60         &     1.22          &      1.24         \\
        50         &     1.13          &      1.14         \\
        45         &     1.09          &      1.10         \\
        40         &     1.06          &      1.07         \\
        35         &     1.04          &      1.04         \\
        30         &     1.02          &      1.02         \\
        25         &     1.01          &      1.01         \\
        20         &     1.00          &      1.00         \\
        15         &     1.00          &      1.00         \\
        10         &     1.00          &      1.00         \\   \hline
        
\end{tabular}                   
\end{table}

\end{appendix}

\end{document}